\documentclass[3p]{elsarticle}

\usepackage{lineno,hyperref}
\usepackage{hyperref}

\usepackage{color}
\usepackage{amssymb, amsmath}
\usepackage{multirow}
\usepackage{url}

\usepackage{algorithm2e}

\usepackage{fancyhdr}
\usepackage{listings}

\journal{Parallel Computing, accepted 3/26/18}

\begin{document}

\thispagestyle{fancy}

\begin{frontmatter}

\title{Parallel Accelerated Vector Similarity Calculations
for Genomics Applications\tnoteref{ornltitlenote}}

\tnotetext[ornltitlenote]{This manuscript has been authored by UT-Battelle,
LLC under Contract No. DE-AC05-00OR22725 with the U.S. Department of
Energy.  The United States Government retains and the publisher, by
accepting the article for publication, acknowledges that the United States
Government retains a non-exclusive, paid-up, irrevocable, world-wide
license to publish or reproduce the published form of this manuscript, or
allow others to do so, for United States Government purposes.  The
Department of Energy will provide public access to these results of
federally sponsored research in accordance with the DOE Public Access Plan
(http://energy.gov/downloads/doe-public-access-plan).
\vskip .05in
Accepted to {\it Parallel Computing}, 3/26/18. \url{https://doi.org/10.1016/j.parco.2018.03.009} \copyright \ 2018. This manuscript version is made available under the CC-BY-NC-ND 4.0 license \url{http://creativecommons.org/licenses/by-nc-nd/4.0/}.
\vskip .05in
}


\author[ornladdress]{Wayne Joubert\corref{mycorrespondingauthor}}
\cortext[mycorrespondingauthor]{Corresponding author}
\ead{joubert@ornl.gov}
\author[ornladdress]{James Nance}
\author[ornladdress,utkaddress]{Deborah Weighill}
\author[ornladdress,utkaddress]{Daniel Jacobson}

\address[ornladdress]{
 Oak Ridge National Laboratory,
 1 Bethel Valley Road,
 Oak Ridge, TN 37831
}

\address[utkaddress]{
The Bredesen Center for Interdisciplinary Research and Graduate Education,
University of Tennessee, Knoxville,
444 Greve Hall, 821 Volunteer Blvd.
Knoxville, TN 37996-3394 }



\begin{abstract}

The surge in availability of genomic data holds promise
for enabling determination of genetic causes of
observed individual traits,
with applications to problems such as discovery
of the genetic roots of phenotypes, be they molecular phenotypes such as gene expression or metabolite concentrations, or complex phenotypes such as diseases.
However, the growing sizes of these datasets
and the quadratic, cubic or higher scaling characteristics
of the relevant algorithms pose a serious computational challenge
necessitating use of leadership scale computing.
In this paper we describe a new approach to performing
vector similarity metrics calculations,
suitable for parallel systems equipped with graphics
processing units (GPUs) or Intel Xeon Phi processors.
Our primary focus is the Proportional Similarity metric
applied to Genome Wide Association Studies (GWAS) and Phenome Wide Association Studies (PheWAS).
We describe the implementation of the algorithms on accelerated processors,
methods used for eliminating redundant calculations due to symmetries,
and techniques for efficient mapping of the calculations 
to many-node parallel systems.
Results are presented demonstrating high per-node performance and
parallel scalability
with rates of more than five quadrillion ($5\times 10^{15}$)
elementwise comparisons achieved per second on the ORNL Titan system.
In a companion paper we describe corresponding techniques applied
to calculations of the Custom Correlation Coefficient
for comparative genomics applications.

\end{abstract}


\begin{keyword}
High performance computing
\sep
parallel algorithms
\sep
NVIDIA~\textsuperscript{\textregistered} GPU
\sep
Intel~\textsuperscript{\textregistered} Xeon Phi
\sep
comparative genomics
\sep
vector similarity metrics
\sep
Proportional Similarity metric
\sep
\MSC[2010]
65Y05
[Computer aspects of numerical algorithms: Parallel computation]
\sep
68W10
[Algorithms: Parallel algorithms]
\end{keyword}

\end{frontmatter}



\section{Introduction}
\label{section:introduction}

The measurement of the similarity of pairs of vectors is a
computation required in many science domains including
chemistry, image processing, linguistics, ecology,
document processing and genomics.
To satisfy domain-specific requirements,
many different similarity measures have been
developed~\cite{cha,weighill2016network}.

The focus of the present study is the use of similarity measures in the analysis of  GWAS and PheWAS results. GWAS analyses involve the statistical association of genetic variants with measured phenotypes~\cite{solovieff2013pleiotropy}. These can be complex phenotypes such as disease states, or molecular phenotypes, such as the concentration of a particular metabolite or the expression level of a particular gene. While GWAS analyses are generally considered to involve testing association of variants with a single or limited number of phenotypes, PheWAS analyses involve testing the association of variants with a large number of different phenotypes~\cite{pendergrass2011use}. The results of GWAS and PheWAS studies can be represented as a matrix of significant associations between variants and phenotypes, and profile vectors variants and phenotypes can be extracted from the rows and columns of this matrix. Pairwise comparisons of these vectors can allow for the discovery of phenotypes affected by similar genetic elements, or of groups of variants which affect similar phenotypes. (For example, see~\cite{bolormaa2014multi}).
These studies, however, are computationally expensive,
insofar as the computational work required for pairwise
comparison grows as the square of the number of vectors.
Even more challenging is the execution of higher-order
studies which consider three or
more vectors at a time---a technique required in order to discover
relationships not discoverable by means of 2-way methods
alone~\cite{3way}---for which the computational complexity is even higher.
In the past, such studies could be performed efficiently
on workstations or small compute clusters.
However, because of the large quantities of data involved,
it is now necessary to employ large-scale high performance computing
to execute scientific campaigns at the largest scales.

This paper describes advances in the development of
algorithms and software to address this need.
We present vector similarity measure calculation techniques for
large datasets run on one of the world's largest compute
systems, scaled to thousands of compute nodes equipped with
GPU accelerators.
The primary contributions of this paper are implementations
of similarity calculation methods which:
1) achieve high absolute performance on GPUs
as a result of careful mapping of calculations to the memory hierarchy and
exploiting of the highly computationally intense BLAS-3-like
structure of the targeted algorithms;
2) use asynchronous internode communication, data transfers
and computations to ameliorate the costs of data motion;
3) strategically arrange the computations to avoid the
potential 2X-6X performance loss factor
arising from the redundant calculations due to symmetry;
4) carefully parallelize the algorithms to enable
near-perfect scalability to thousands of compute nodes on
leadership-class systems.

In this paper we focus on the 2-way and 3-way variants of
the Proportional Similarity metric, also known as the
Czekanowski metric~\cite{czek,3way},
using an approach that is generalizable to other metrics.
In the companion paper~\cite{companionccc}
we describe corresponding work on
Custom Correlation Coefficient (CCC)~\cite{ccc} calculations
with applications to comparative genomics.


Improving computational throughput for performing comparisons
between pairs, triples or larger subsets of
a set of vectors
has been the focus of significant recent work
centering around the use of parallelism, accelerated GPU or Intel Xeon Phi
processing, or both.
A broad overview of epistasis detection in comparative genomics
including computational issues pertaining to parallelism and GPU acceleration
is given in \cite{wei}.
The GBOOST code, discussed in
\cite{gboost},
is a gene-gene interaction code for 2-way
studies optimized for single GPUs using encoding of gene data into bit
strings with avoidance of redundant computations;
\cite{gwisfi}
describes GWISFI, a single-GPU code for 2-way GWAS calculations.
\cite{gonzalez}
develops a UPC++ code for gene-gene interaction studies
for small numbers of GPUs and Intel Phi processors
exploiting vector hardware and hardware population count instructions.
\cite{gonzalez2}
considers 3-way interactions on a node with 4 GPUs.
\cite{solomonik}
develops parallel tensor computation methods,
structurally similar to 3-way metrics computations, with
particular attention to avoiding redundant computations;
however, the work does not consider GPUs or shaping of the
computational regions to accommodate processors with long vector lengths.
\cite{pande}
discusses similarity metric calculations for chemical informatics
applications
on single GPUs using space filling curve methods and hardware
population count instructions;
it recognizes the correspondence of these calculations to
BLAS-3 matrix-matrix product computations and pays close
attention to optimizing memory accesses.
\cite{cloud} considers 2-way studies on compute clouds using MapReduce
on conventional CPUs.
\cite{pawsey}
adapts existing packages to perform 2-way CPU and GPU
studies and 3-way CPU studies on as many as 200 cores in
parallel.
\cite{goudey}
performs k-way GWAS studies for arbitrary k with consideration of
load balancing and elimination of redundancies
on a 4096-node IBM Blue Gene/Q system;
results for a single GPU are also presented.
\cite{weeks}
performs 2-way analyses on up to 126 nodes of the Intel
Phi-based Stampede system (cf. \cite{weeks2}).
\cite{stanzione}
considers 2-way computations on thousands of compute cores
with good scalability and good absolute performance on
conventional CPUs.
Finally, recent work in \cite{fb}
considers $k$-selection similarity search methods with applications
to image data with results for small numbers of GPUs;
that work however focuses primarily on the $k$-selection problem for
nonexhaustive inexact similarity search,
a different problem from what is considered here.

The present study is to our knowledge the first work
bringing together all the required ingredients for
high performance 2-way and
3-way comparative genomics studies on modern
leadership-class systems:
use of accelerated processors at high absolute performance;
optimization of calculations for use with complex memory hierarchies;
elimination of redundant computations;
algorithm and code design to minimize costs of I/O;
and careful arrangement of communications for near-ideal
scaling to many thousands of compute nodes.

The remainder of this paper is structured as follows.
After describing the 2-way and 3-way Proportional Similarity
metrics in Section~\ref{section:czekanowski},
we describe the techniques used to map these methods to GPUs
and other manycore accelerated processors in
Section~\ref{section:gpus}.
Then we describe the parallelization techniques applied to these methods
in Section~\ref{section:parallel},
followed by implementation details in
Section~\ref{section:implementation}.
Computational results on the 27 petaflop
Oak Ridge National Laboratory (ORNL) Cray XK7 Titan system
are presented in Section~\ref{section:results},
and conclusions are given in
Section~\ref{section:conclusions}.


\def\minprod{\circ_{min}}

\section{The Proportional Similarity metric}
\label{section:czekanowski}

\subsection{The 2-way metric}

We assume a set of $n_v$ vectors of length $n_f$ elements
$\{v_i\}_{i=1}^{n_v}$ with $v_i\in\mathbb{R}^{n_f}$ and
$v_i=\{v_{i,q}\}$.
In practice, $v_{i,q}\geq 0$.
Then the 2-way Proportional Similarity metric for two vectors $v_i$ and
$v_j$ is $c_2(v_i,v_j)=2 n_2(v_i,v_j)/d_2(v_i,v_j)$
where
$d_2(v_i,v_j)={\sum_q v_{i,q} + v_{j,q}}
={\sum_q v_{i,q}}
+{\sum_q v_{j,q}}$ and
$n_2(v_i,v_j)={\sum_q v_{i,q} \minprod v_{j,q}}$.
Here $a\minprod b={\rm min}(a,b)$,
which we define here (for reasons to be described shortly)
as the ``min-product,''
is simply the function returning the minimum value of two scalars.

Due to symmetry of the $c_2()$ function with respect to its arguments,
computing the metric for all pairs of distinct vectors
requires computing only ${n_v}({n_v}-1)/2$ distinct values, for example,
$\{c_2(v_i,v_j)\}_{i=1}^{n_v}{}_{j=i+1}^{n_v}$.
Note that to perform this computation,
computing the denominators $d_2()$
requires only $({n_f}-1){n_v} = O({n_f}{n_v})$ scalar adds,
used to compute ${\sum_q v_{i,q}}_i$.
The numerators $n_2()$ however require $({n_f}-1){n_v}({n_v}-1)/2$ adds and
${n_f}{n_v}({n_v}-1)/2$ min-product computations\footnote{
In principle the min-product complexity
could be reduced to $O(n_f n_v \log(n_v))$ by presorting each row of $V$;
however, it is unclear whether this would accrue a performance gain
on modern cache-based processors,
and furthermore the number of floating point additions would remain unchanged.
},
thus having complexity
$O({n_f}{n_v}^2)$.

With each vector $v$ interpreted as a distribution,
the Proportional Similarity metric $c_2(u,v)$ of two vectors
is large if $u$ and $v$
have a peak at the same vector entry locations
$q$ for many index values $q$ (see~\cite{Vegelius1986}).


\subsection{The 3-way metric}

The 3-way Proportional Similarity metric is defined by
$$c_3(v_i,v_j,v_k)={3\over 2}{n_3(v_i,v_j,v_k)\over d_3(v_i,v_j,v_k)}$$
where
$d_3(v_i,v_j,v_k)={\sum_q v_{i,q} + v_{j,q} + v_{k,q}}$ and
$$n_3(v_i,v_j,v_k)=n_2(v_i,v_j) + n_2(v_i,v_k) + n_2(v_j,v_k) -
n'_3(v_i,v_j,v_k)
\eqno{(1)} $$
where
$n'_3(v_i,v_j,v_k)={\sum_q v_{i,q} \minprod v_{j,q} \minprod v_{k,q}}$.

The function $c_3()$ is symmetric in its arguments,
thus only ${n_v}({n_v}-1)({n_v}-2)/6$ distinct values need be computed, for example,
$\{c_3(v_i,v_j,v_k)\}_{i=1}^{n_v}{}_{j=i+1}^{n_v}{}_{k=j+1}^{n_v}$.
As before, the denominators $d_3()$ require
$({n_f}-1){n_v} = O({n_f}{n_v})$ adds.
The numerators $n_3()$ require one computation of each of the three
$n_2()$ values, computed as described above and
used via table lookup for the first three terms of $n_3()$,
and additionally calculation of the last term $n'_3()$
requiring
$({n_f}-1){n_v}({n_v}-1)({n_v}-2)/6$
adds and
$2{n_f}{n_v}({n_v}-1)({n_v}-2)/6$
computations with the min-product operation, thus complexity
$O({n_f}{n_v}^3)$.

The value of the 3-way Proportional Similarity metric is large when two
of the three vectors have large values at corresponding
entries $q$ and even larger if all three vectors have large
values at matching entries $q$.


\subsection{Related metrics}

The Sorenson metric is identical to the Proportional Similarity metric
for the special case when $v_{i,q}\in\{0, 1\}$ for all $i$,
$q$.
Though Sorenson metric values can be computed using methods for computing
the Proportional Similarity metric,
the computation can be made much faster
on most processors in general use by
representing vector entries as bits packed into words and
operated upon using binary arithmetic,
based on the coincidence of the min-product and the bitwise
logical AND operations for this case.


\section{Mapping to manycore processors}
\label{section:gpus}

\subsection{The 2-way metric}

Since for large $n_f$ and $n_v$ the computation of the numerators $n_2()$ by
far dominates the runtime, it is adequate to accelerate this computation only
on the GPU;
in the present work, all other computations are performed on the CPU.

Let $V$ represent the matrix of column vectors
$V=[ v_1 v_2 \cdots v_{n_v} ]$.
For arbitrary matrices $A=\{a_{ij}\}$ and $B=\{b_{ij}\}$, define
$A\minprod B$ by $(A\minprod B)_{ij} = \sum_k a_{ik}
\minprod b_{kj}$.
It is manifest that the desired numerators
can be specified as a subset of the entries of
the matrix $M = V^T \minprod V$,
for example, the strict upper triangular entries.

Note that the operation $A\minprod B$ has identical
computational pattern to the standard matrix-matrix product operation
$A\cdot B$ for a general full matrix (GEMM), the former
being defined by simply replacing the
standard scalar multiplication operation $a\cdot b$
of the GEMM with the $a\minprod b$ operation.
This GEMM operation of the BLAS-3 standard \cite{blas3}
is one of the most highly optimized
kernels in high performance computing and is supported by
many heavily optimized libraries, typically yielding near-peak
floating point operation rates for targeted processors.
The approach we take here therefore is to optimize $V^T \minprod V$
performance
by adapting existing highly optimized linear algebra software to perform
this operation.
In what follows, we will refer to the
$A\minprod B$ computation as a ``modified GEMM'' or mGEMM operation.

On modern architectures it is of paramount importance that
computations be optimized to the memory hierarchy, including
registers, caches and main memory.
For dense linear algebra, the complex coding effort to
optimize algorithms to the memory hierarchy has already been done in
the form of mathematical libraries such as
the open source MAGMA library~\cite{magma}, which we use here.
Though we do not pursue the topic here, it is likely that
libraries optimized to other processor architectures, such as
PLASMA~\cite{plasma}, BLIS~\cite{blis} and OpenBLAS
\cite{openblas},
would provide similar opportunities for adapting to the
mGEMM operation, in this case for conventional processors and Intel Xeon Phi.

A potential performance concern of this approach is
that only half of the entries of $M=V^T \minprod V$ are
required, whereas all standard implementations of GEMM
compute every entry of this matrix, resulting in potential
performance loss of a factor of two.
A possible remedy is to break the matrices into smaller
blocks and skip computation of lower triangular blocks of
$M$; however, this would still result in some performance
loss since GEMM computations are most efficient at large matrix sizes.
These issues are not a concern here, however;
Our primary focus is the case of many compute nodes,
in which case most of the time is spent in off-diagonal
block computations of the form $W^T\minprod V$ for distinct
$W$ and $V$ (see below), for which there are no wasted
computations.
Thus the performance impact of this issue is minor.

Standard GEMMs attain high performance by spending most of
the time in fused multiply-add (FMA) operations
$c \leftarrow c + a \cdot b$, which are typically optimized in
hardware to execute in a single clock cycle.
The operation of taking the minimum of two values, however, is not so
well-optimized as FMA and may in fact require a branch in execution flow,
depending on the implementation.
This puts a theoretical limit on the performance of
Proportional Similarity metric computations that is somewhat less than
that of matrix-matrix product computations.
To maximize performance, here we use hardware intrinsics for
taking the minimum of two values, available on recent NVIDIA GPU
hardware.

Attaining high performance on a GPU-accelerated node
requires overlapping transfers to and from the GPU with
computations on the GPU.
In this case, the mGEMM computation can be broken into
blocks whose computation is overlapped with transfers.
A more effective approach however is to overlap the
computations $W^T\minprod V$ of the  multiple off-diagonal blocks
in the parallel case with both GPU transfers and
node-to-node communications in a pipelined fashion using
double buffering.
This is efficient since the case of many nodes requires many
mGEMMs pertaining to off-diagonal blocks to be computed, as
will be described below.

It is apparent that the approach described here is applicable
to other vector similarity metrics that are likewise based on
the accumulation of the results of a scalar operation applied to
corresponding pairs of vector elements.


\subsection{The 3-way metric}

The calculation of 3-way metrics requires calculation of
2-way numerators as described above as well as denominators;
the most expensive part for problems
of significant size however is
calculation of the 3-way term involving $n'_3()$.

This computation could be described as ``BLAS-4-like''
based on its computational pattern and complexity.
However, since the BLAS-3-like 2-way computation
already are able to approach theoretical processor peak performance,
for our purposes it will be sufficient to
decompose the 3-way computation into a sequence of 2-way
computations, these in turn executed
as described earlier, without relying per se on the
BLAS-4-like local structure of the problem.

Define the matrix $X_j$ such that
$(X_j)_{ik} = (V)_{ij} \minprod (V)_{ik}$.
The columns of $X_j$ are simply the elementwise min-products
of the columns of $V$ with the single column $v_j$.
Then let $B_j = X_j^T \minprod V$.
Observe $(B_j)_{ik} = n'_3(v_i,v_j,v_k)$.
Thus $B_j$ can readily be computed on GPUs using the 2-way
method described earlier.

Due to symmetries it is necessary to compute only the
bottom $n_v-j+1$ rows of $B_j$.
Even after this optimization, some redundant calculations
still remain; however, as with the
2-way case, when many GPU-enabled nodes are used most computations
involve off-node blocks for which the relevant matrix entries
are distinct and thus redundancy is not present.
The use of only a single compute node is not the focus of
performance optimization here,
though in our experience even with some performance losses
the methods described here may run many times faster than
conventional methods at low node counts.

Since the 3-way calculation requires a sequence of GPU
kernel calls corresponding to 2-way operations for
the $B_j$ matrices described above, the
calculation lends itself naturally to overlapping the
transfers of data to and from the GPU with the computations on the
GPU in a pipelined fashion using double buffering,
thus minimizing the overhead of transfers.
Furthermore, many such kernel calls can be performed
without the need for intervening off-node communication
in the parallel case,
this being an added performance benefit resulting from the BLAS-4-like
nature of the computation.


\section{Multi-node parallelism}
\label{section:parallel}

\subsection{The 2-way case}

We assume a large set of compute nodes, each equipped with local
memory and one GPU, with interconnect programmable via MPI;
this case is easily generalizable to multiple GPUs per node
by assuming each GPU is associated with a single MPI rank on the node.

\begin{figure}[ht]
\centering
\includegraphics*[height=1.5in]{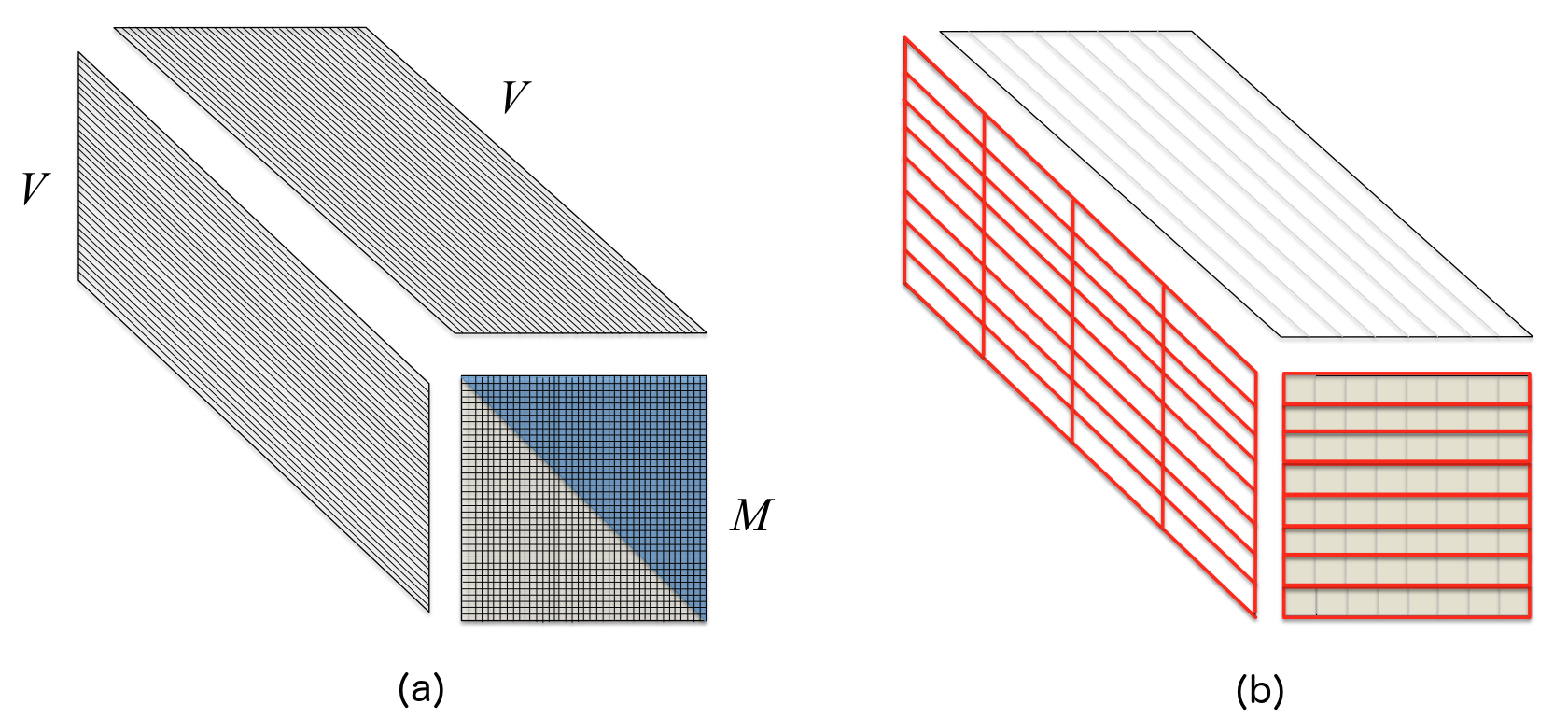}
\caption{(a): Computational pattern of 2-way metric calculation;
(b) Data decomposition for 2-way metric calculation}
\label{fig:2way-comp-pattern-decomp}
\end{figure}

For the 2-way case, the primary aim is to compute the
2-D square matrix $M$ of results corresponding to the numerators
$n_2()$ described earlier, approximately half of which are unique,
describable for example by a triangular set of values
(Figure~\ref{fig:2way-comp-pattern-decomp}(a)).

Considering the set of column vectors in the matrix $V$,
at least two possible
axes of parallelism exist.  First, each vector can be split
into multiple pieces and assigned to compute nodes
(parallelism in vector elements: partitioning of rows of $V$).
Second, a subset of the
vectors can be assigned to each compute node (parallelism in
vector number: partitioning of columns of $V$),
with each node storing
the rows of $M$ corresponding to the owned vectors
(Figure~\ref{fig:2way-comp-pattern-decomp}(b)).
In the present work we allow both axes of parallelism in
arbitrary combination.
Let $n_{pf}$ and $n_{pv}$ denote the number of nodes in the
decomposition along the vector element and vector number
axes, respectively, with vector elements and counts per node
represented by $n_{fp}$ and $n_{vp}$.

Parallelizing along the vector elements axis
requires a reduction of locally summed values along this axis
to accumulate results.
This typically requires $log(n_{pf})$ communication steps,
each with nodes communicating $n_{vp}=n_v/n_{pv}$ values per node.
Relying on this axis alone for parallelism is problematic
for several reasons:
(1) for larger values of $n_{pf}$, the
logarithmic growth in communication cost with respect to node count will
eventually dominate, even if communication is hidden under computation;
(2) the problem sizes targeted in practice by comparative
genomics problems sometimes result in small values for the vector length per node
$n_{fp} = n_f/n_{pf}$ for large $n_{pf}$, leading to low efficiencies.
For these reasons, we only consider modest amounts of parallelism
along this axis.
We might consider additional
performance improvement from use of asynchronous reduction
operations here; we will not pursue this in the present work.

Decomposing along the vector number axis necessitates an all-to-all
communication.
Though such operations are in general expensive, advantage can be
taken here of the special structure of the problem.
We formulate the computation as a sequence of parallel steps.
At step $i\geq 0$, each node computes numerator values for
the comparison of its own vectors with
vectors stored on the node that is $i$ nodes away in the upward
direction, with wraparound if needed.
This process is pipelined and double-buffered,
both between nodes and between each node's CPU and GPU,
so that at any parallel step, computations of numerators on the GPU,
GPU transfers in each direction, communications, and computations of
denominators and quotients on the CPU are all overlapped.
Due to the computational intensity of the mGEMM,
for sufficiently large per-node problem sizes,
GPU computations will fully overlap all other operations.

\begin{figure}[ht]
\centering
\includegraphics*[height=1.0in]{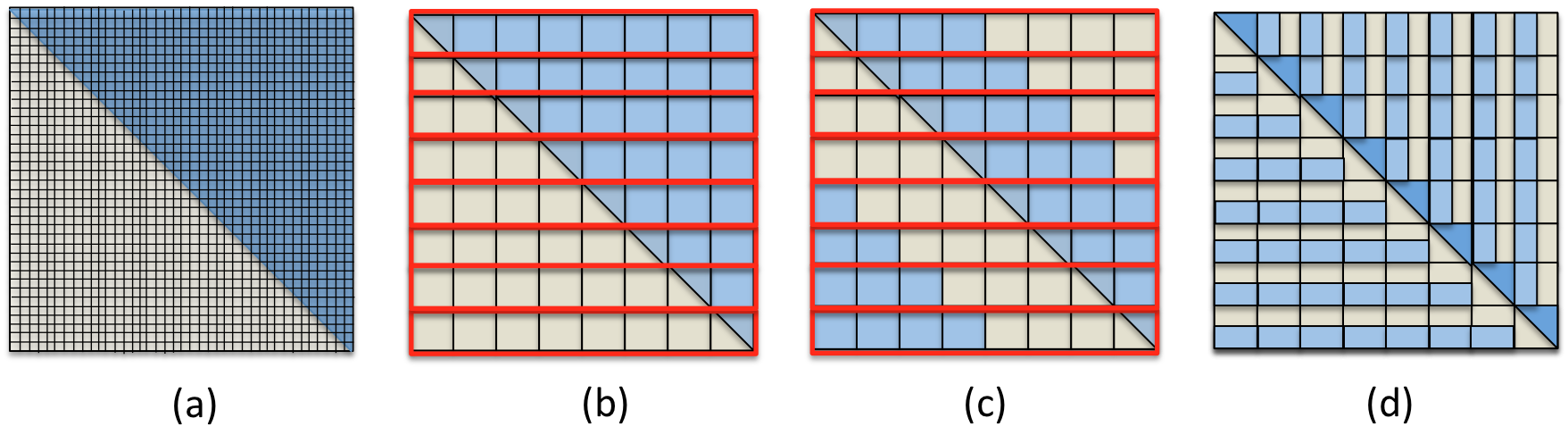}
\caption{Load balancing for 2-way case.
(a) na\"{\i}ve computation of upper triangular entries of result;
(b) load imbalance for na\"{\i}ve approach;
(c) load balanced method using block circulant structure of
computed values;
(d) alternative load balancing strategy}
\label{fig:2way-load-balancing}
\end{figure}

This approach must be implemented so as to avoid
computation of unneeded, redundant values.
Using a na\"{\i}ve approach based on computing only the upper
triangular matrix entries would result in load imbalance:
some block rows assigned to compute nodes would have
much less work than others and be idle for part of the computation;
see Figure~\ref{fig:2way-load-balancing}(b).
The solution adopted here is to compute results associated with
a block circulant-structured subset of the matrix blocks
(Figure~\ref{fig:2way-load-balancing}(c));
this subset has the properties that all unique values are
represented exactly once and also each block row has the same
amount of work, resulting in load balance.

Past experience, e.g., with the ScaLAPACK project~\cite{scalapack},
has shown that distributed dense linear algebra based on
only a 1-D decomposition of the target matrix may
not provide sufficient
parallelism\footnote{The parallelization strategies used here and
in ScaLAPACK are related to the concepts of sharding,
vertical partitioning and replication
used in distributed database management.}.
To generalize the above approach,
we thus define an additional parallelism axis.
A parameter $n_{pr}$ is chosen,
and the computations of the
blocks corresponding to a block row of the matrix $M$ are
distributed to the $n_{pr}$ compute nodes in round-robin fashion.
We thus have total internode parallelism of
$n_p = n_{pf} n_{pv} n_{pr}$ across $n_p$ compute nodes.
The specific decomposition is selected to tune for
optimal performance for the targeted
case.

The pseudocode in Algorithm~\ref{alg:2way} demonstrates the computation of
metrics using the vector elements $V_{f,v}$ stored
on node $(p_f, p_v, p_r)$.
This code, unlike the actual code, assumes for simplicity certain
divisibility conditions of variables and also
does not overlap communications, GPU transfers and computations.

\begin{algorithm}
  \caption{2-way metrics computation}
  \label{alg:2way}

  Put $V_{f,v}$ to GPU

  \For{$\Delta{}p_{ji} = 0$ to $\lfloor n_{pv} / 2 \rfloor$} {

    \If{$\mod(\Delta{}p_{ji}, n_{pr}) = p_r$} {

        $p_{v_s} = mod(p_v - \Delta{}p_{ji}, n_{pv})$; $p_{v_r} = mod(p_v + \Delta{}p_{ji}, n_{pv})$

        Send $V_{f,v}$ to $(p_f, p_{v_s}, p_r)$; Receive $V_{f,v_r}$ from $(p_f, p_{v_r}, p_r)$

        Put $V_{f,v_r}$ to GPU; Compute $N_{f,v,v_r} \leftarrow V_{f,v_r}^T \minprod V_{f,v}$ on GPU; Get $N_{f,v,v_r}$ from GPU

        Compute $N_{v,v_r}$ from $N_{f,v,v_r}$ by reduction

        Compute denominators; Compute metrics
    }
  }
\end{algorithm}


\subsection{The 3-way case}

\begin{figure}[ht]
\centering
\includegraphics*[height=1.5in]{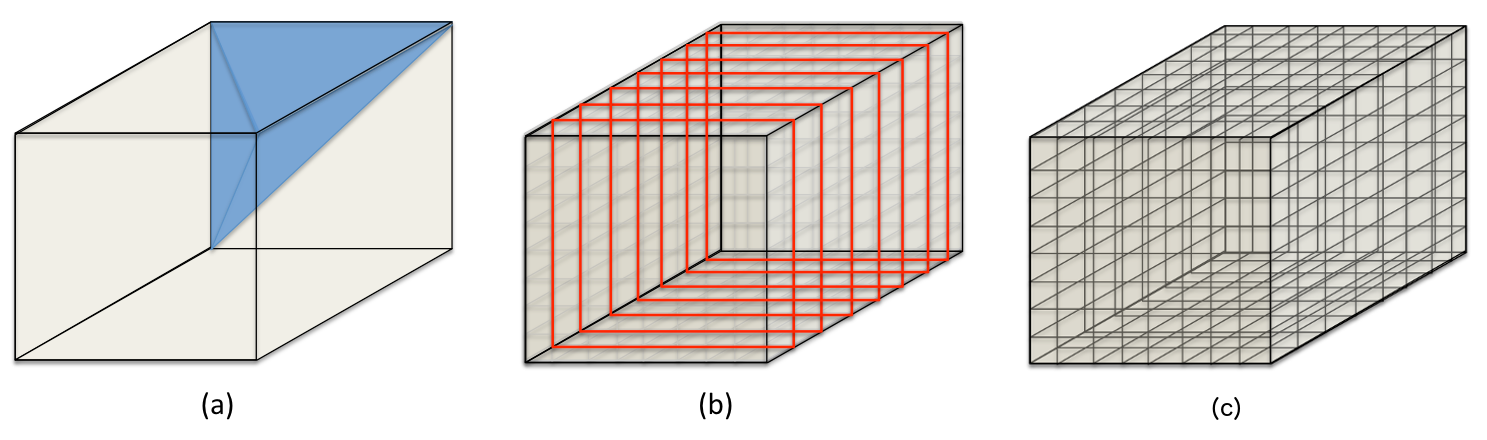}
\caption{
(a) Domain of results for 3-way metric with a representative
subregion of unique values;
(b) the results domain resulting from parallel decomposition
across vector number;
(c) induced decomposition of domain into blocks}
\label{fig:3way-decomp-3}
\end{figure}

For the 3-way case, a 3-D cube of results must be computed,
approximately 1/6 of which are unique.
A tetrahedral region represents one possible selection of unique values
(Figure~\ref{fig:3way-decomp-3}(a)).

As previously, we decompose along the vector element axis
and the vector number axis.
For the vector element axis, we use a nonblocking reduction
operation, since a sequence of mGEMM operations is readily
available to overlap this reduction.

For the vector number decomposition, analogously to the
2-way case we partition the set of vectors,
inducing a 1-D decomposition of the cube of results
into small cube-shaped blocks arranged
into ``slabs,'' each owned by the respective compute node
(Figure~\ref{fig:3way-decomp-3}(b)).
To accomplish the all-to-all communication,
a doubly nested loop is required for any given node to
obtain the vectors required to compute its slab of results,
analogously to the singly nested loop of the 2-way case.
In particular, compute node $i$, owning vector partition $i$,
must compute comparisons with vector partitions $j$ and $k$,
within the nested parallel step loops over all $j$ and $k$.
Each parallel step in turn entails a sequence of mGEMM operations
pipelined with overlapping of transfers to and from the GPU.
This inner pipeline is thus itself a component of an outer pipeline
of parallel steps which overlaps the required communications
of vector blocks with the mGEMM calculation pipeline.

As with the 2-way case, the redundant calculations
for the 3-way case must be eliminated, to avoid potential 6X
performance loss.
It is unclear that the 2-way approach generalizes;
we thus seek an alternative.
To motivate the approach,
Figure~\ref{fig:2way-load-balancing}(d) shows for the 2-way case
an alternative selection of unique values satisfying the property
that not only every row but also every block has approximately
the same number of computed results.
Furthermore one can see by reflection across the main
diagonal that all unique values are represented exactly once.

\begin{figure}[ht]
\centering
\includegraphics*[height=1.5in]{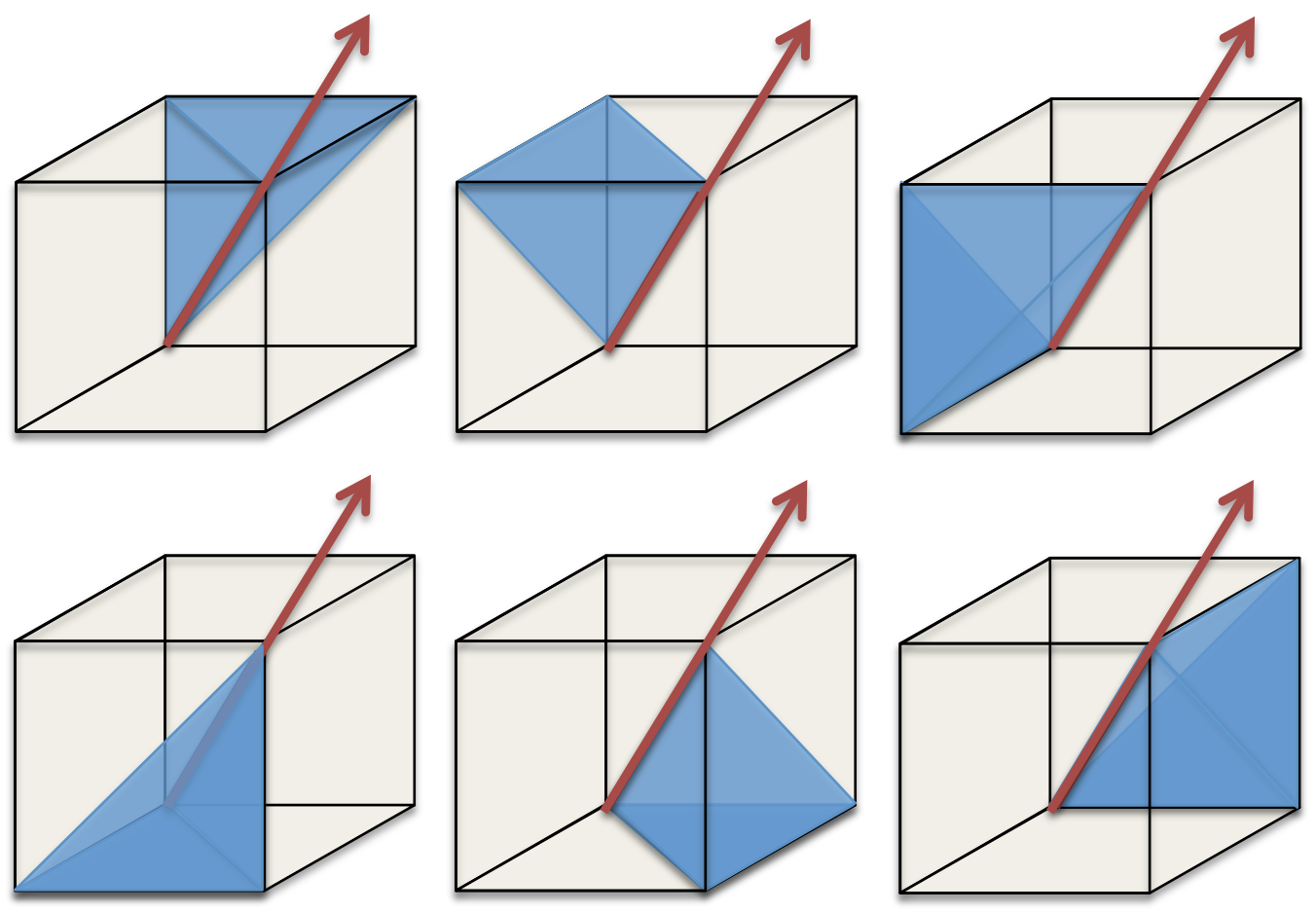}
\caption{Decomposition of the results domain into disjoint tetrahedra}
\label{fig:3way-tetrahedra}
\end{figure}

To generalize to the 3-way case we consider a partitioning
of the results cube into six tetrahedra, each sharing an
edge with the main diagonal of the cube
(Figure~\ref{fig:3way-tetrahedra}).
Onto this partitioning is superimposed the implicit tiling
into blocks induced by the parallel decomposition
(Figure~\ref{fig:3way-decomp-3}(c)).

Following the approach of
Figure~\ref{fig:2way-load-balancing}(d),
we now consider three types of blocks:
blocks on the main diagonal of the cube (``diagonal edge blocks''),
blocks on an interior face of a tetrahedron (``face blocks''),
and the remaining blocks in the volume of the cube (``volume blocks'').
To define the algorithm, as a first step we select for each
block a strategic subset of 1/6 of the values to compute
so as to cover all unique values in a load balanced fashion.
We then adjust this approach to improve performance.

\begin{figure}[ht]
\centering
\includegraphics*[height=1.3in]{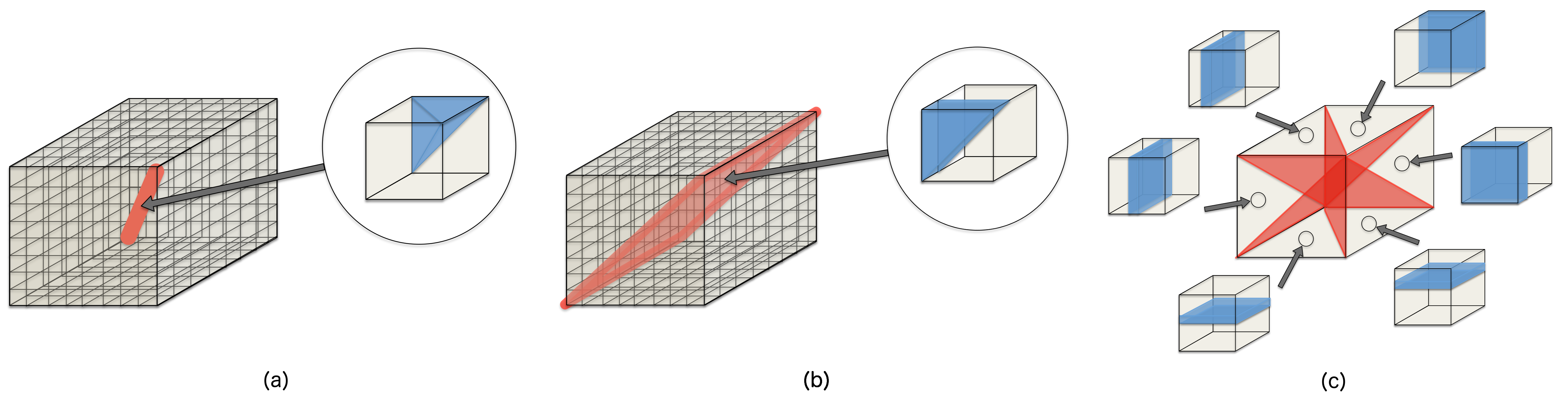}
\caption{Compute region for main diagonal edge blocks (a), face blocks (b) and volume blocks (3)}
\label{fig:3way-block-part123}
\end{figure}

The subset of values to compute for each block is selected as follows.
For diagonal edge blocks, we select a small tetrahedron of result values
(Figure~\ref{fig:3way-block-part123}(a)).
For face blocks, one of three possible
1/3-height triangular prisms is selected,
with orientation and positioning depending
on the location in the cube;
an example is given in Figure~\ref{fig:3way-block-part123}(b).
Finally, for volume blocks, one of six 1/6-thickness slices
of the block is selected,
with placement and orientation based on location in the
domain (Figure~\ref{fig:3way-block-part123}(c)).
By a sequence of folding and reflection operations
it can be shown that these combined
selections are equivalent to
the single large tetrahedron of values in
Figure~\ref{fig:3way-decomp-3}(a).

For significant numbers of nodes, most of the time
is spent computing volume blocks.
To compute a volume block, the shorter dimension of the
slice is used as the GPU pipeline direction, so that the
mGEMMs have maximal size and run at high efficiency.

As with the 2-way case, it is possible to extract additional
parallelism.
To accomplish this, the blocks computed by the doubly nested
loop are round-robin distributed 
to a set of $n_{pr}$ compute nodes
assigned to the slab.
Again we have total internode parallelism of
$n_p = n_{pf} n_{pv} n_{pr}$ across $n_p$
compute nodes.

In the actual implementation the above scheme is modified slightly.
First, to simplify scheduling, the three diagonal planes
of face blocks are folded into a single plane,
so that the 1/3-height triangular prisms are replaced with a
smaller set of full-height prisms;
because of the structure of the computation,
this change does not introduce load balancing concerns.
Second, though the method described assigns an equal number of result
values to each block, the work per block is not equal
since volume blocks execute much more efficiently than
diagonal edge and face blocks, resulting in load imbalance
for high values of $n_{pr}$.
To resolve this, we additionally divide diagonal edge blocks and
face blocks each into six slices and distribute these slices
along the $n_{pr}$ axis.
As a result, each slab of the domain now has
$6 + 6 (n_{pv}-1) + (n_{pv}-1)(n_{pv}-2) = (n_{pv}+1)(n_{pv}+2)$
slices to compute.
This eliminates the load balancing problem while
introducing a slight load
imbalance factor of $n_{pv}^2 / (n_{pv}+1)(n_{pv}+2)$
which becomes insignificant as the node count $n_{pv}$ is made large.

An additional modification is made to improve efficiency.
For the 3-way case, the number of metrics to be computed
and stored can be enormous due to the algorithm's scaling complexity.
This has two tangible adverse impacts.
First, nodal memory requirements constrain the problem size
per node to be very small, thus reducing mGEMM efficiency.
Second, the wallclock time to compute the entire complement
of metrics at once may exceed the queue limit policies at some
computing facilities.
To remedy this, for the 3-way case we implement a
staging capability: the entire run campaign for a set of metrics
can be decomposed into $n_{st}$ stages, with only a single stage
of results computed and stored at a time.
This is implemented by dividing the GPU pipeline described
earlier into $n_{st}$
parts and computing and storing the metrics for only one
part at a time.
This allows mGEMM sizes and thus efficiencies to be
increased substantially.

As described above, the 3-way computation is composed of an
inner GPU computation pipeline wrapped in an outer
communication pipeline.
Algorithm~\ref{alg:3way} shows pseudocode demonstrating the communication
pipeline for computation of
metrics using the vector elements $V_{f,v}$ stored
on node $(p_{f}, p_{v}, p_{r})$.
The pseudocode in Algorithm~\ref{alg:3wayblock}
describes the GPU pipeline computing
the metrics associated with $(V_{f,v}, V_{f,v_{r,j}}, V_{f,v_{r,k}})$,
a component of Algorithm~\ref{alg:3way}.
In both cases a simplified form is shown without the asynchronous behavior
of the actual implementation.

\begin{algorithm}
  \caption{3-way metrics communication pipeline}
  \label{alg:3way}

  Put $V_{f,v}$ to GPU; block/slice counter $s_b \leftarrow 0$

  \For{$s = 0$ to $5$} {
    \If{$mod(s_b, n_{pr}) = p_r$} {

        Compute $V_{f,v}$ denominators

        Compute metrics for block using $(V_{f,v}$, $V_{f,v}$, $V_{f,v})$
    }
    $s_b \leftarrow s_b + 1$
  }

  \For{ $s = 0$ to $5$} {

    \For{$\Delta{}p_{ji} = 1$ to $n_{pv} - 1$} {

      \If{$mod(s_b, n_{pr}) = p_r$} {

          $p_{v_s} = mod(p_v - \Delta{}p_{ji}, n_{pv})$; $p_{v_r} = mod(p_v + \Delta{}p_{ji}, n_{pv})$

          Send $V_{f,v}$ to $(p_f, p_{v_s}, p_r)$; Receive $V_{f,v_r}$ from $(p_f, p_{v_r}, p_r)$

          Put $V_{f,v_r}$ to GPU

          Compute $V_{f,v_r}$ denominators

          Compute metrics for block using $(V_{f,v}, V_{f,v_r}, V_{f,v_r})$
      }

      $s_b \leftarrow s_b + 1$
    }
  }

  \For{$\Delta{}p_{ki} = 1$  to  $n_{pv} - 1$} {

    $p_{v_{s,k}} = mod(p_v - \Delta{}p_{ki}, n_{pv})$; $p_{v_{r,k}} = mod(p_v + \Delta{}p_{ki}, n_{pv})$

    Send $V_{f,v}$ to $(p_f, p_{v_{s,k}}, p_r)$;  Receive $V_{f,v_{r,k}}$ from $(p_f, p_{v_{r,k}}, p_r)$

    Put $ V_{f,v_{r,k}}$  to  GPU

    Compute $V_{f,v_{r,k}}$ denominators

    \For{$\Delta{}p_{ji} = 1$ to $n_{pv} - 1$} {

      \If{$mod(s_b, n_{pr}) = p_r$ and $\Delta{}p_{ji} \not= \Delta{}p_{ki}$} {

          $p_{v_{s,j}} = mod(p_v - \Delta{}p_{ji}, n_{pv})$; $p_{v_{r,j}} = mod(p_v + \Delta{}p_{ji}, n_{pv})$

          Send $V_{f,v}$ to $(p_f, p_{v_{s,j}}, p_r)$;  Receive $V_{f,v_{r,j}}$ from $(p_f, p_{v_{r,j}}, p_r)$

          Put $V_{f,v_{r,j}}$ to GPU

          Compute $V_{f,v_{r,j}}$ denominators

          Compute metrics for block using $(V_{f,v}, V_{f,v_{r,j}}, V_{f,v_{r,k}})$
      }

      $s_b \leftarrow s_b + 1$
    }
  }
\end{algorithm}

\begin{algorithm}
  \caption{3-way metrics GPU pipeline for slice $s$ in block and stage $s_t$ using $(V_{f,v}, V_{f,v_{r,j}}, V_{f,v_{r,k}})$}
  \label{alg:3wayblock}

  Compute $N_{f,v_{r,j},v} \leftarrow V_{f,v_{r,j}}^T \minprod V_{f,v}$ on GPU; Get \ $N_{f,v_{r,j},v}$ from GPU; Reduce to $N_{v_{r,j},v}$

  Compute $N_{f,v_{r,k},v} \leftarrow V_{f,v_{r,k}}^T \minprod V_{f,v}$ on GPU; Get $N_{f,v_{r,k},v}$ from GPU; Reduce to $N_{v_{r,k},v}$

  Compute $N_{f,v_{r,j},v_{r,k}} \leftarrow V_{f,v_{r,j}}^T \minprod V_{f,v_{r,k}}$ on GPU; Get $N_{f,v_{r,j},v_{r,k}}$ from GPU; Reduce to $N_{v_{r,j},v_{r,k}}$

  $j_{min} \leftarrow \lfloor (s_t + n_{st}s)(n_v/n_{pv}) / (6n_{st}) \rfloor$; $j_{max} \leftarrow \lfloor (s_t + 1 + n_{st}s)(n_v/n_{pv}) / (6n_{st}) \rfloor$

  \For{$j = j_{min}$ to $j_{max} - 1$} {

    Compute $X_j$ columns from $V_{f,v}$, $e_j^T V_{f,v_{r,j}}$; Put $X_j$ columns to GPU

    Compute $B_j$ rows; Get $B_j$ rows from GPU; reduce

    Compute metrics
  }
\end{algorithm}


\section{Implementation}
\label{section:implementation}

The algorithms described here are implemented in
the CoMet parallel genomics code.
This code is written in C++, compiles with the GNU compiler suite
and depends on MPI, CUDA and the modified versions of the MAGMA library.
GNU Make and CMake are used for build management, and googletest
is used for unit testing.
The clang-format source code tool from the clang compiler package is used for
source code formatting, and Git is used for repository management.

OpenMP CPU threading is used to accelerate the parts of the
computation that are not ported to the GPU
by mapping execution to multiple CPU
cores on the node;
when possible, the CPU work is also hidden under the
asynchronously launched GPU kernels to improve performance.

For making comparisons, each method has a reference (CPU-only) version,
a (possibly optimized) CPU version, and a GPU version.
A set of synthetic reference test cases is implemented for
testing, designed to give the exact same bit-for-bit result
for all code versions and for all parallel decompositions.
Two types of synthetic problem are implemented:
a version for which each vector entry is set to a randomized value,
and a second version with randomized placement of entries specifically
chosen so that the correctness of every result value can be
verified analytically.
A checksum feature using extended precision integer
arithmetic computes a bit-for-bit exact checksum of
computed results to check for errors when using
synthetic inputs.

The code can be compiled under single or double precision.
The single precision version requires less compute time and is
adequate if the vector lengths are sufficiently short
and the number of digits of precision required in the result is
sufficiently low.

To modify MAGMA as needed for the algorithms, it is
necessary to modify the two files in the MAGMA package
{\tt{}magmablas/gemm\_stencil.cuh} and
{\tt{}magmablas/gemm\_stencil\_defs.h}.
In particular, the macro definition for ``{\tt{}fma}'' defining the
fused multiply-add must be changed to make use of the
min-product operation.


\section{Computational results}
\label{section:results}

\subsection{Overview}

Experiments are performed on the ORNL Titan Cray XK7
system.
Titan is composed of 18,688 compute nodes each equipped with
an AMD Interlagos 16 core CPU and an NVIDIA Kepler K20X GPU
connected via a PCIe-2 bus.
The K20X GPU has peak single/double precision flop rate of
3,935/1,311 GF and peak memory bandwidth of 250 GB/s. 
Each node contains 32 GB main memory and 6 GB GPU memory.

The software versions used are
Cray OS version 5.2.82,
Cray Programming Environment 2.5.5,
GCC  4.9.3,
MAGMA 1.6.2
and
CUDA toolkit 7.5.18-1.0502.10743.2.1.
For large node counts, it is in some cases necessary to set the
environment variable {\tt{}APRUN\_BALANCED\_INJECTION}
to values such as 63 or 33 to avoid throttling of the
communication network resulting from the
algorithms' communication patterns
and causing performance loss.

The primary use of the code is to solve very large
problems not previously solvable; thus weak scaling behavior,
for which the work per node is kept roughly constant as
compute node count is increased, is the primary focus.

GPU-enabled runs are executed
with one MPI rank and one GPU per Titan node.
Reported execution times do not include I/O and setup costs.
The source code execution path for the algorithm
is identical independent of
the actual values contained in the input vectors;
thus we expect performance for the synthetic datasets used
here to be essentially identical to performance with actual
genomics data.


\subsection{Single GPU kernel performance}

We first validate that the modified MAGMA kernel
has comparable performance to the true GEMM operation.
Table~\ref{tab:1gpu} shows results for a sample case
with $n_v=10,240$ vectors of length $n_f=12,288$ elements
run on a single node.
Kernel times are taken from the CUDA Profiler
and include kernel time only, without transfer or CPU times.
GEMM achievable performance figures are taken from~\cite{K20X}.
Though the CUDA intrinsics {\tt{}fmin} and {\tt{}fminf} are used in
the production code, timings using the C ternary conditional
operator are additionally included for comparison.

It is apparent from the results that the performance of the
modified GEMM kernel is a large fraction of achievable peak
GEMM rate.
There is an expected degradation of performance from using
a {\tt{}fmin} or {\tt{}fminf} hardware intrinsic
combined with a scalar addition
instead of FMA,
insofar as FMA can execute in a single clock cycle
unlike the {\tt{}fmin} or {\tt{}fminf} operation combined with an addition.
Furthermore, the MAGMA standard GEMM on which the methods are
based is not as fast as the cuBLAS GEMM.
This is because
the MAGMA GEMM is specifically optimized for smaller matrix sizes
required by other MAGMA operations rather than the large
sizes in focus here.
Though the performance is high, a topic of future
study is to improve the performance of this kernel, which is the
ultimate performance determiner of the algorithm.


\begin{table}
\begin{center}
\caption{Kernel times in seconds for single GPU case}
\label{tab:1gpu}
\vspace{.05in}
\begin{tabular}{ccc}
                                  & single    & double    \\
                                  & precision & precision \\
\hline 
mGEMM, c += a \textless b ? a : b & 3.056     & 7.222     \\
mGEMM, CUDA intrinsic fminf/fmin  & 2.602     & 6.484     \\
GEMM, MAGMA                       & 2.097     & 4.179     \\
GEMM, cuBLAS                      & 1.035     & 2.410     \\
GEMM achievable peak              & 0.889     & 2.112     \\
GEMM theoretical peak             & 0.655     & 1.966     \\
\hline 
\end{tabular}
\end{center}
\end{table}



\subsection{Performance model}

It is desirable to model algorithm performance in order to
evaluate expected performance and also to give guidance
regarding selection of tuning parameters.
We assume here that mGEMM sizes are large enough to hide
communications, GPU data transfers and CPU computations.

For the 2-way case, we define $\ell$, the ``load,''
to denote the number of blocks assigned to each node.
Then the execution time of the algorithm is estimated by
$$ t = t_C + t_{T,V} + \ell \cdot t_G + t_{T,M} + t_{CPU} , $$
where $t_C$ is the time for communicating $n_{fp} n_{vp}$ vector elements
per node for a parallel step,
$t_{T,V}$ the time to transfer $n_{fp} n_{vp}$ vector elements
to the GPU for a step,
$t_{T,M}$ the time to transfer $n_{vp}^2$ metrics values
from the GPU per step,
$t_{CPU}$ the time for denominator and quotient calculations
per step
and $t_G$ the time for an mGEMM computation.
The non-mGEMM times are included here to account for pipeline startup and drain,
assuming the mGEMMs perfectly overlap the other operations.
It is evident that maximizing $\ell$ (by limiting $n_{pr}$)
makes it possible to approach peak mGEMM performance.
mGEMM rates are determined empirically; the
goal is to make the matrix dimensions $n_{fp}$ and
$n_{vp}$ as large as possible to maximize mGEMM efficiency.
This suggests for a given problem it is desirable to reduce $n_{pv}$ and
$n_{pf}$ until CPU or GPU memory is filled.

For the 3-way case, we again define the load $\ell$, here
representing the number of block slices computed by a node.
Each slice is computed by a GPU pipeline of
$(n_{vp}/6)/n_{st}$ mGEMM steps.
The execution time of the algorithm is estimated by
$$ t = t_C + t_{T,V} + \ell \cdot
 [ (3+(n_{vp}/6)/n_{st})t_G + 3 t_{T,V} + 4t_{T,M} + t_{CPU} ] . $$
mGEMM performance is approached by increasing $\ell$, and
$n_{vp}$ and decreasing $n_{st}$, subject to memory
constraints.
The value of $n_{st}$ should be kept small to minimize the
impact of the three 2-way metrics calculations required for
each slice.
As with the 2-way case, $n_{fp}$ and $n_{vp}$ should be
maximized in order to maximize mGEMM performance.


\subsection{GPU vs.\ CPU performance comparison}

The following results compare performance with and
without use of GPUs, to evaluate the performance advantage
afforded by use of GPUs.
These results use 20,000 fields on 32 compute nodes
decomposed via $n_{pv}=32$ using double precision and
200,000 (2-way case) or 6,144 (3-way case) vectors.
Results are shown in Table~\ref{tab:gpucpu}.
The GPU version performs better than CPU by factors
of 41X (2-way case) and 27X (3-way case).
These figures should be compared against the roughly 10X
peak flop rate ratio and 5X peak memory bandwidth ratio between
GPU and CPU for each Titan node.
Since the CPU code version used here
is a reasonable implementation but not as heavily optimized
as the GPU version,
these results should not be interpreted to reflect
the comparative best achievable performance
by a highly optimized code for each case.

\begin{table}
\begin{center}
\caption{GPU vs. CPU runtime comparisons in seconds}
\label{tab:gpucpu}
\vspace{.05in}
\begin{tabular}{cccc}
num way & GPU & CPU & ratio \\
\hline 
2  &  76.8 & 3149.9 & 41.0 \\
3  & 371.3 & 10067  & 27.1 \\
\end{tabular}
\end{center}
\end{table}


\subsection{Strong scaling results}

\begin{figure}[ht]
\centering
\includegraphics*[height=2.0in]{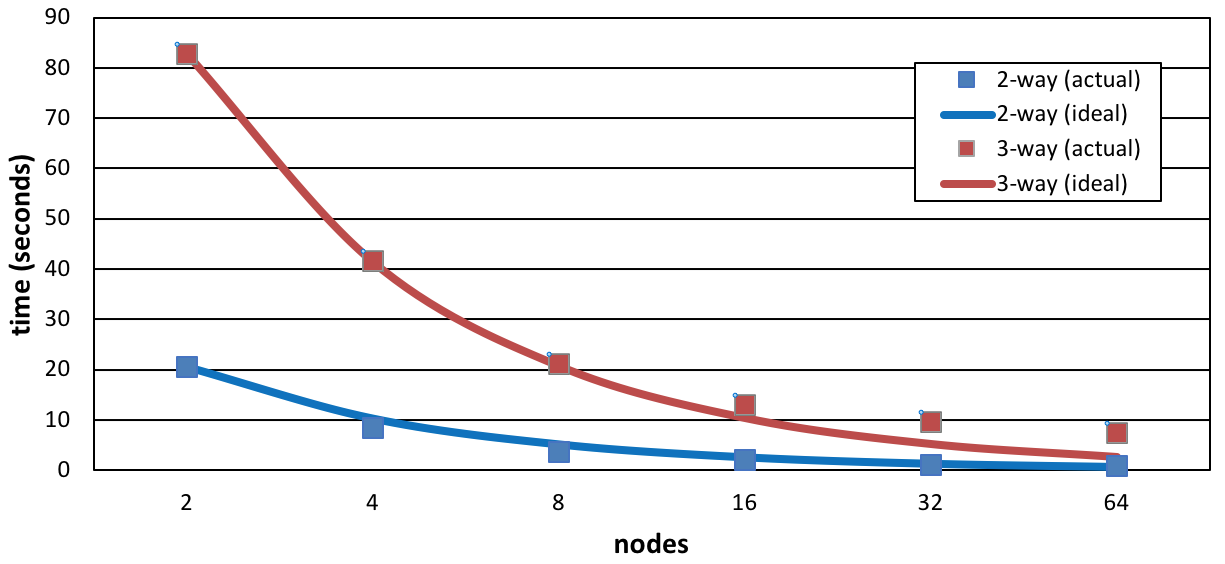}
\caption{Proportional Similarity metric 2-way and 3-way double precision strong scaling runtimes}
\label{fig:czek-double-strongscaling}
\end{figure}

Though the primary application of this work is weak scaling
regimes, we present strong scaling results for completeness.
Figure~\ref{fig:czek-double-strongscaling}
shows double precision results for a fixed problem
with 20,000 fields and
16,384 (2-way case) or 1,544 (3-way case) vectors.
The problem is run on 2 to 64 compute nodes with various
processor decompositions; the best case for each node count
is shown.
The parallel efficiencies at 64 nodes relative to 2 nodes
are 79\% (2-way) and 34\% (3-way).
The 3-way case is less efficient here, since the large number of
metrics to be computed constrains the problem size to be
smaller than what would be typically run in practice.
Efficiency could be substantially increased
by computing only a single stage at a time out of many compute stages,
enabling larger matrix sizes;
doing this here however would not be a
strong scaling test in the commonly understood sense, since the the subset of
elements selected for a compute stage is not independent of
the processor decomposition, thus a slightly different problem would
be solved in each case.

\subsection{2-way weak scaling results}

Executing a problem on $n_p$ nodes
requires selecting tuning parameters
$n_{pf}$, $n_{pv}$ and $n_{pr}$ satisfying $n_{pf} n_{pv} n_{pr} = n_p$.
We set $n_{pf}=1$ for these experiments since actual
datasets these cases are intended to emulate have modest
sizes for $n_f$; nonetheless, experiments show that the code
has very good weak scaling behavior along this axis for modest $n_{pf}$.

For fixed $n_{pv}$, setting $n_{pr}=\lceil n_{pv} / 2 + 1 \rceil$
assigns a single block to each node.
We thus set $n_{pr} = \lceil \lceil n_{pv} / 2 + 1 \rceil / \ell \rceil$
where $\ell$ is the load.

Experiments are performed for up to 17,472 of Titan's 18,688 compute
nodes, or 93.5\% of the system.
The double precision case uses $n_f=5,000$ elements per vector,
$n_{vp}=10,240$ vectors per node and load $\ell=13$.

Though the BLAS-3 nature of the mGEMM makes it possible to
asynchronously hide communication costs for large problems,
the 2-way Proportional Similarity communication costs are still
challenging, in this case for example requiring messages
of nearly 1/2 GB size.
To maximize communication performance,
we execute the 2-way runs in dedicated system mode
with environment variables
{\tt{}APRUN\_BALANCED\_INJECTION=96},
{\tt{}ARMCI\_DMAPP\_LOCK\_ON\_GET=1} and
{\tt{}ARMCI\_DMAPP\_LOCK\_ON\_PUT=1}.
Furthermore, following an approach previously used
for optimizing parallel transpose operations~\cite{ptrans},
we apply a randomly generated mapping of the problem to compute
nodes, using the {\tt{}MPICH\_RANK\_REORDER\_METHOD}
environment variable and a specifying a random reordering
by input file.
Failure to use these two techniques at scale often resulted in
network throttling events leading to irregular runtimes and decreased
application performance by a factor of 2X or more.

\begin{figure}[ht]
\centering
\includegraphics*[height=1.5in]{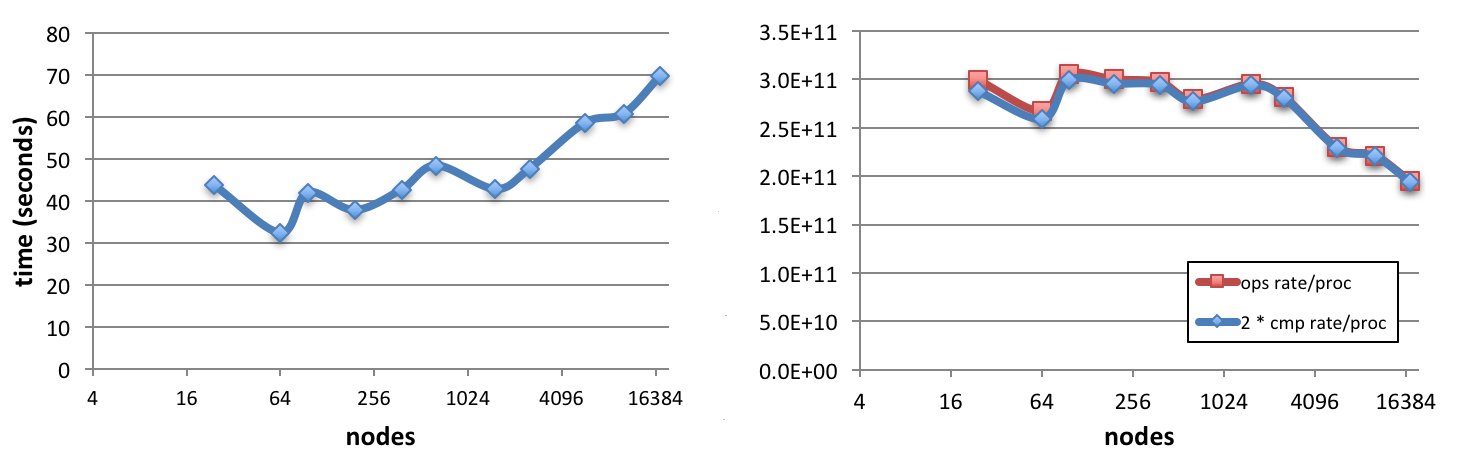}
\caption{Proportional Similarity metric 2-way double precision weak scaling. Left: time to solution. Right: operations (add, multiply, fmin) per second, and corresponding number of unique elementwise comparisons per second per node.}
\label{fig:czek-double-2way-results}
\end{figure}

Figure~\ref{fig:czek-double-2way-results} shows weak
scaling results.
Timings are shown in the left graph.
Though there is some loss of performance
due to communication costs
as the node count is increased,
the performance loss is a mere 37\% as the node count is
increased by nearly three orders of magnitude.
The right graph shows rate of operations per node,
where scalar addition, scalar multiplication
and scalar minimum are each counted as one operation.
Since the comparison of two vector elements
requires a scalar minimum and a scalar add,
the value of twice the comparison rate is also shown.
The values nearly match, except at low node counts
for which the unneeded computations for the main diagonal
block cause slight loss of performance.
The implied operation rate for large cases
derived from Table~\ref{tab:1gpu} is roughly 398 GOps/sec per node
(1 GOp = $10^9$ operations), to which the rates in
Figure~\ref{fig:czek-double-2way-results} should be compared.
The maximum comparison rate for the largest case is 1.70
petacomparisons per second; see Table~\ref{tab:czek-2way-comparisons-max}.

In practice, the use of a random permutation of nodes might
be expected to result in some performance variability based
on the specific permutation used;
in production, if needed one could in principle save a ``good''
permutation for a given node count and problem setup
for subsequent reuse.
However, we believe code modifications are possible which would make it
unnecessary to use a random
permutation of nodes: by an adjustment of the code
it should
be possible to recast the communication pattern as a
nearest neighbor communication and then use known methods
to map the communication pattern optimally to the network,
see, e.g. \cite{gampi}; this will be a topic of future study.
In any case, as with other parallel applications, optimizing communications
in a multiuser environment is challenging insofar as the
network bandwidth is shared by other users and furthermore it is not always
possible for a user to reserve a communication-optimal subset of nodes
for job execution.

\begin{figure}[ht]
\centering
\includegraphics*[height=1.5in]{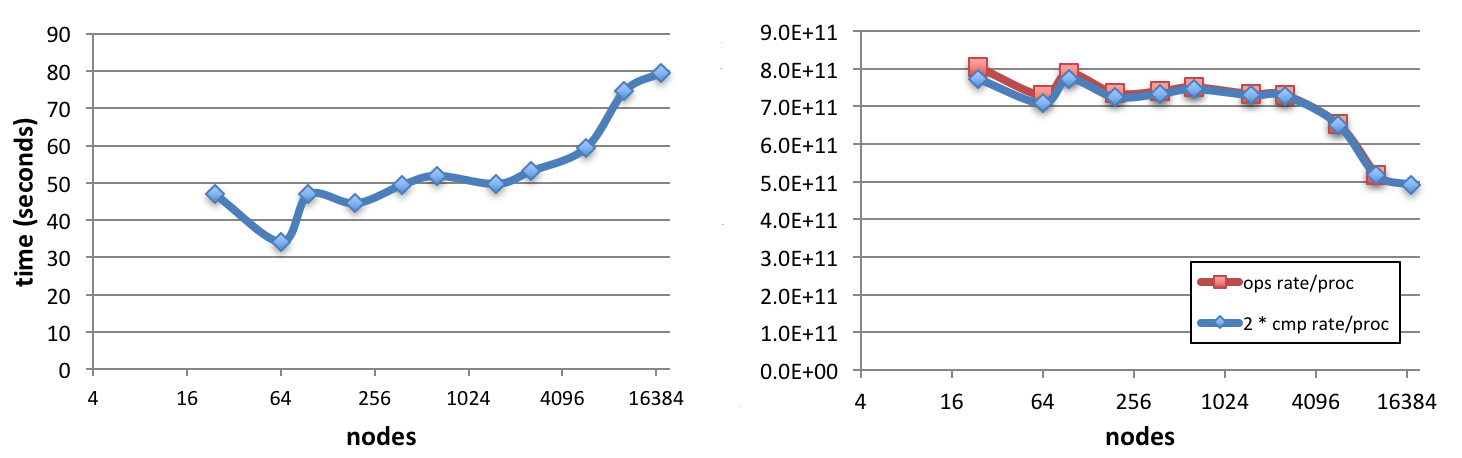}
\caption{Proportional Similarity metric 2-way single precision weak scaling. Left: time to solution. Right: operations (add, multiply, fmin) per second, and corresponding number of unique elementwise comparisons per second per node.}
\label{fig:czek-single-2way-results}
\end{figure}

The single precision test cases use $n_f=10,000$ elements per vector,
$n_{vp}=12,288$ vectors per node and load $\ell=13$.
Figure~\ref{fig:czek-single-2way-results} shows the weak
scaling results.
Results are qualitatively similar to the double precision case,
with rate over twice as
fast, owing to the use of single precision.
The performance loss is only 41\% as the node count is
increased by nearly three orders of magnitude.
The implied operation rate for large cases
derived from Table~\ref{tab:1gpu} is roughly 991 GOps/sec per node,
to which these rates should be compared.
The maximum comparison rate for the largest case is 4.29
petacomparisons per second.

\begin{table}
\begin{center}
\caption{Maximum performance, 2-way Proportional Similarity metric}
\label{tab:czek-2way-comparisons-max}
\vspace{.05in}
\begin{tabular}{ccc}
method & operations & comparisons \\
       & per second & per second \\
\hline 
double precision & $ 3.40\times 10^{15}$ & $1.70\times 10^{15}$ \\
single precision & $ 8.59\times 10^{15}$ & $4.29\times 10^{15}$ \\
\hline 
\end{tabular}
\end{center}
\end{table}


\subsection{3-way weak scaling results}

For the 3-way case we must select $n_{pf}$, $n_{pv}$,
$n_{pr}$ and additionally the stage count $n_{st}$.
For load $\ell$ we set
$n_{pr} = \lceil ( n_{pv} + 1 ) ( n_{pv} + 2 ) / \ell \rceil$.
The stage count $n_{st}$ for best efficiency should be set to divide
$n_{vp}/6$ evenly, for $n_{vp}=n_v / n_{pv}$ vectors per node.

\begin{figure}[ht]
\centering
\includegraphics*[height=1.5in]{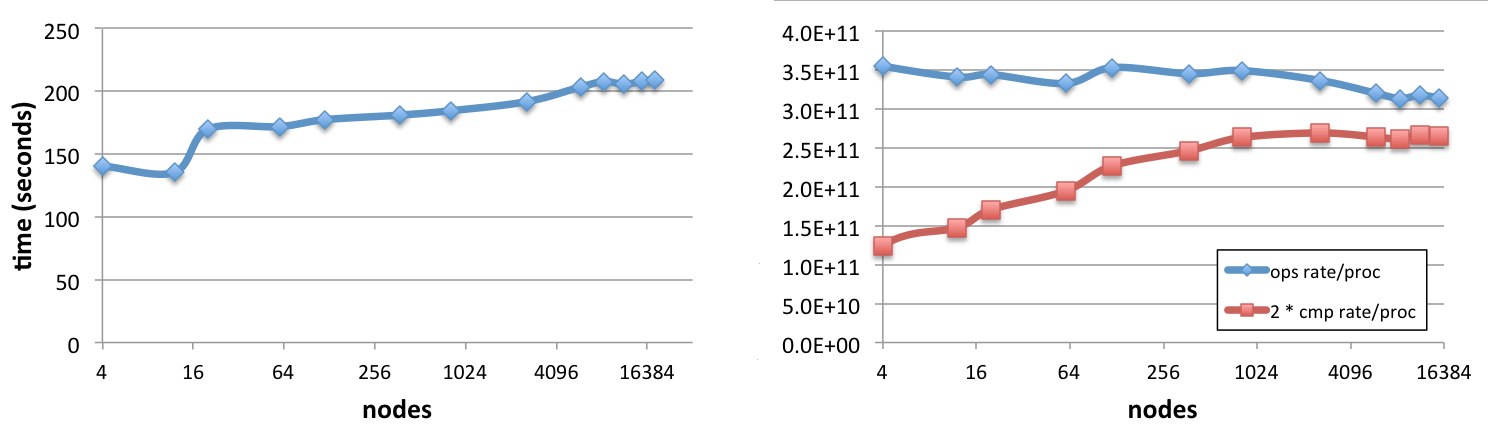}
\caption{Proportional Similarity metric 3-way double precision weak scaling. Left: time to solution. Right: operations (add, multiply, fmin) per second, and corresponding number of unique elementwise comparisons per second per node.}
\label{fig:czek-double-3way-results}
\end{figure}

The test runs are executed on up to 
18,424 of Titan's 18,688 compute nodes, or 98.6\% of the system.
Figure~\ref{fig:czek-double-3way-results} shows results
for the double precision case, with $n_f=20,000$ elements per vector and
$n_{vp}=2,880$ vectors per node, computing the final stage
of $n_{st}=16$ stages (pipeline depth 30), with load $\ell=6$.
The left graph shows very good weak scaling behavior,
with some anomalies at low node counts for which per-node
performance is less efficient.
The right graph shows the average operations per node
primarily dominated by mGEMM costs.
The rate is maintained above 300 GOps per node up
to the highest node counts, a figure to be compared to the
double precision GEMM rate of up to 398 GOps implied by Table~\ref{tab:1gpu}.
This is compared in the figure to twice the comparison rate
per node.
At lower node counts there is some disparity between the
two, primarily caused by the load balancing issues described
earlier, but at high node counts these two quantities
approximate each other, indicating that overheads are low;
the remaining disparity is mainly the result of the required 2-way
computations at pipeline startup,
which are counted here as part of the 3-way
comparison operation.

\begin{figure}[ht]
\centering
\includegraphics*[height=1.5in]{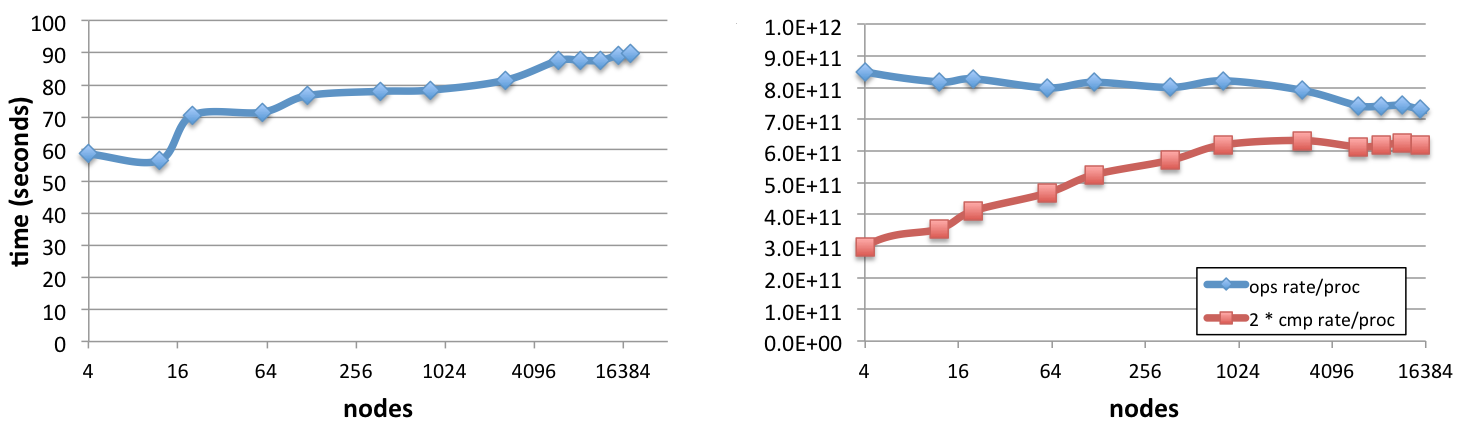}
\caption{Proportional Similarity metric 3-way single precision weak scaling. Left: time to solution. Right: operations (add, multiply, fmin) per second, and corresponding number of unique elementwise comparisons per second per node.}
\label{fig:czek-single-3way-results}
\end{figure}

Figure~\ref{fig:czek-single-3way-results} shows corresponding results for
the single precision case, with $n_f=20,000$ elements per vector and
$n_{vp}=2,880$ vectors per node, computing the final stage
of $n_{st}=16$ stages, with load $\ell=6$.
The results are qualitatively similar to the double
precision case, with over 2X higher performance due to increases in
instruction rate and memory bandwidth.

Table~\ref{tab:czek-3way-comparisons-max} shows the maximum
operation and comparison rates attained at the highest node counts, up to
2.44 petacomparisons per second (double precision)
or 5.70 petacomparisons per second (single precision).

\begin{table}
\begin{center}
\caption{Maximum performance, 3-way Proportional Similarity metric}
\label{tab:czek-3way-comparisons-max}
\vspace{.05in}
\begin{tabular}{ccc}
method & operations & comparisons \\
       & per second & per second \\
\hline 
double precision & $ 5.75\times 10^{15}$ & $2.44\times 10^{15}$ \\
single precision & $13.40\times 10^{15}$ & $5.70\times 10^{15}$ \\
\hline 
\end{tabular}
\end{center}
\end{table}


\subsection{Results for a realistic sample problem}

We now show results for data corresponding to an actual
(i.e., nonsynthetic) problem.
The input data is from a Phenome Wide Association dataset of poplar
metabolites,
containing all of the single nucleotide polymorphisms (SNPs)
that have a significant genome wide association study (GWAS)
association to one or more metabolites measured across a
GWAS population of poplar trees. GWAS was performed using EMMAX~\cite{emmax}.
This case has $n_v=189,625$ vectors of length $n_f=385$.

The runs are performed in single precision; this is satisfactory
on account of
the short vector lengths and modest accuracy requirements.
Due to the comparatively short vector length for this
dataset, it is expected that
the mGEMM operations will run at less than optimal efficiency;
therefore, 
for comparison, a synthetic problem is also run with the same
dimensions and settings except that the vector length
$n_f$ is chosen larger.  All cases set $n_{pf}=1$.
The input data is stored in a single column-major binary file;
each compute node reads the required portion of this file.
The output is written as one file per node with each metric
value written as a single unsigned byte value storing
roughly 2-1/2 significant figures, an adequate size for this study;
all metrics are written with no thresholding.
No indexing information need be written explicitly
since this information can be computed formulaically
offline.
No further efforts are made to optimize I/O performance.

The 2-way method is run using $n_p=n_{pv}=30$ Titan nodes.
The 3-way method is run with $n_{pv}=30$ and $n_{pr}=496$
on $n_p=n_{pv}\cdot n_{pr}=14,880$ Titan nodes; in this case
only the last stage of $n_{st}=220$ stages is computed.
Ouput is measured only for the 2-way cases.
Results are given in Table~\ref{tab:sample-problem-timings}.
Performance is good for the sample case but substantially
better if the vector length is longer, due to the improved
mGEMM performance.

It is apparent that the high speed of the metrics calculation
itself requires that
careful attention be given to workflow design
for scientific campaigns, specifically the handling of
the large volume of generated metrics data.
The trend in high performance computing system hardware
is increasing compute node speed, while I/O bandwidths are growing at
a much slower rate;
burst buffers will increase this rate, but their storage capacities
will be limited.
Techniques for performing in situ analysis on the nodes before
writing out results will reduce the data burden.
Methods to threshold, downsample and compress data and optimize the
data writing process will be important for efficient
performance.
Any requirements to use global methods for determining
the thresholding values or other output tuning parameters will be
problematic when using staging, since in this case not all of the results
are available in memory at any given time to compute the needed
parameters.
The staging option for the 3-way case however will allow favorable opportunities
for overlapping of computation and output: a stage can be computed
at the same time that
asynchronously the previous stage is written to disk.
This strategy will require some care however,
since main memory bandwidth is required
for the output as well as for the data transfers
to and from the GPU during compute and also for communication;
these competing operations will need to be carefully
scheduled in order to avoid resource contention.

\begin{table}
\begin{center}
\caption{Sample problem timings in seconds (unoptimized I/O)}
\label{tab:sample-problem-timings}
\vspace{.05in}
\begin{tabular}{cccccc}
num way & $n_f$  & input & metrics comp & output & comparison \\
        &        & time  & time         & time   & rate / node \\
\hline 
 2      & 385    & .06    &  1.85        & 24.78  & 125e9 \\
 2      & 20,000 & ---    & 28.86        & 24.79  & 415e9 \\
 3      & 385    & 13.89  & 15.38        & ---    &  54e9 \\
 3      & 5,000  & ---    & 33.37        & ---    & 321e9 \\

\hline 
\end{tabular}
\end{center}
\end{table}


\subsection{Comparison with other work}

We now compare against results reported in the
literature as described in Section~\ref{section:introduction}.
We consider the most relevant comparable work
we are aware of as of this writing.
Table~\ref{tab:comparisons} shows comparisons per second
for alternative methods and implementations.
Unfortunately, it is difficult to make rigorous comparisons
when architectures and algorithms are significantly different and
source code for the methods is not available or is not easily
ported and tuned to a single common architecture for comparison.
Furthermore, most reported results in the literature
for vector similarity for genomics applications represent input values
as bits rather than full floating point values.
This makes comparisons tenuous insofar as manipulating many binary
values packed into a word is disproportionately much faster than
operating on floating point numbers.
With these caveats in mind we report element comparisons per second
in Table~\ref{tab:comparisons}, whether an element is a single bit,
a string of 2 or 3 bits, or a floating point number.
Also as a rough measure of this performance normalized against
hardware capability,
a normalized performance ratio is calculated by normalizing
the (absolute) comparison rate
against the floating point rate of the respective hardware.
For newer hardware, the double precision peak rate is used;
for older GPUs with weak or nonexistent double precision support,
single precision is used.

We consider first the raw number of comparisons per second.
Table~\ref{tab:comparisons} demonstrates that
the CoMet code used in the present work
gives several orders of magnitude higher absolute
performance, as measured in comparisons per second,
far beyond the demonstrated capability of any other code.
Furthermore, the 3-way method's normalized performance ratio
is as high as 5X higher than that of any other 3-way implementation shown.

The normalized performance ratio of CoMet is within the general range of
performance ratios of the other methods.
Also, the CoMet 2-way single precision method's rate
is within about a factor of four
of the best performing 2-way GWAS method.
This should be considered a strong result,
insofar as the bitwise GWAS methods operate on 2-bit or 3-bit values,
disproportionately easier since they are
less than one-tenth the size of the 32 bit floating point
values operated on by the current code.
Also, pointing back to Table~\ref{tab:1gpu}, CoMet
code performance is already approaching peak available performance
of the underlying hardware; thus we believe that
in a fair apples-to-apples comparison the present code would rank well.
Finally, the performance of \cite{pande} is exceptionally high
since the codes operate on single bit values
and the CPU code exploits native high performance features of
that processor model.

\begin{table}
\begin{center}
\caption{Comparisons to related work}
\label{tab:comparisons}
\vspace{.05in}
\begin{tabular}{cccccccc}
code & problem & node config & nodes & GFlop rate & cmp/sec         & norm \\
     &         &             & used  &            & ($\times 10^9$) & perf \\
\hline 
\cite{pande}            & 2-way 1-bit& 1 Intel Core i7-920& 1 & 42.56 DP       & 222     &5.216 \\
GBOOST\cite{gboost}     & 2-way GWAS & 1 NVIDIA GTX 285   & 1 & 1062.72 SP     & 64.08   & .060 \\
GWISFI\cite{gwisfi}     & 2-way GWAS & 1 NVIDIA GTX 470   & 1 & 1088.6 SP      & 767     & .705 \\
\cite{goudey}           & 2-way GWAS & 1 NVIDIA GTX 470   & 1 & 1088.6 SP      & 649     & .596 \\
\cite{pande}            & 2-way 1-bit& 1 NVIDIA GTX 480   & 1 & 1345 SP        & 1185    & .881 \\
\cite{goudey}           & 2-way GWAS & IBM Blue Gene/Q    & 4096 & 839e3 DP    & 2520    & .003 \\
epiSNP\cite{weeks}      & 2-way GWAS & 2 Intel Xeon Phi SE10P & 126 & 271e3 DP & 1593    & .006 \\
\cite{gonzalez}         & 2-way GWAS & 2 NVIDIA K20m +    & 1 & 3360.56 DP     & 1053    & .313 \\
                        &            & 1 Intel Xeon Phi 5110P & &              &         &      \\
CoMet                   & 2-way PS SP& 1 NVIDIA K20X      & 17472 & 25.3e6 DP  & 4.290e6 & .169 \\
CoMet                   & 2-way PS DP& 1 NVIDIA K20X      & 17472 & 25.3e6 DP  & 1.700e6 & .067 \\
\hline 
GPU3SNP\cite{gonzalez2} & 3-way GWAS & 4 NVIDIA GTX Titan & 1 & 6000 DP        & 264.7   & .044 \\
CoMet                   & 3-way PS SP& 1 NVIDIA K20X      & 18424 & 26.7e6 DP  & 5.700e6 & .213 \\
CoMet                   & 3-way PS DP& 1 NVIDIA K20X      & 18424 & 26.7e6 DP  & 2.440e6 & .091 \\
\end{tabular}
\end{center}
\end{table}


\section{Conclusions}
\label{section:conclusions}

We have defined a new set of algorithm implementations
for computing 2-way and 3-way
vector comparison metrics on leadership class systems,
showing performance of up to
five quadrillion vector element comparisons per second.
To our knowledge this is the first computation of its kind
ever performed at this scale,
demonstrating the capability to 
perform simulations that were until recently considered
far beyond what is possible,
enabling new kinds of science in
GWAS and PheWAS to be done.

Future work will include investigation of
additional code performance
improvements---for example, improved mGEMM
performance and lessening of overheads by additional code
tuning.
We anticipate the methods described here will be
directly
portable to alternative processor architectures such as
Intel Xeon Phi.
We will also examine implementation of higher order comparison
methods.
Use of these methods will require careful engineering of
computational science
workflows in such a way that the entire workflow is
well optimized for scientific discovery.


\section*{Acknowledgments}

This research used resources of the Oak Ridge Leadership Computing
Facility at the Oak Ridge National Laboratory, which is supported by
the Office of Science of the U.S. Department of Energy under Contract
No. DE-AC05-00OR22725.

This research was funded by the BioEnergy Science Center (BESC) at the Oak Ridge National Laboratory (contract DE-PS02-06ER64304). BESC is a U.S. Department of Energy Bioenergy Research Center supported by the Office of Biological and Environmental Research in the DOE Office of Science. This research was also supported by the Plant-Microbe Interfaces Scientific Focus Area (http://pmi.ornl.gov) in the Genomic Science Program, the Office of Biological and Environmental Research (BER) in the U.S. Department of Energy Office of Science. Oak Ridge National Laboratory is managed by UT-Battelle, LLC, for the US DOE under contract DE-AC05-00OR22725. 

We would like to acknowledge the following people: Timothy Tschaplinski, Priya
 Ranjan, Nan Zhao and Madhavi Martin for the metabolomics data; Nancy Engle,
David Weston, Ryan Aug, KC Cushman, Lee Gunter and Sara Jawdy
for metabolomics sample collection; Carissa Bleker and
Piet Jones for performing the GWAS; Gerald Tuskan, Wellington Muchero
and the DOE Joint Genome Institute (JGI) for sequencing the
\textit{Populus} genotypes and generating the processed the SNP data.


\section*{References}

\bibliography{genomics_czek_paper}


\end{document}